\documentclass[11pt]{article}
\pdfoutput=1
\usepackage[utf8]{inputenc}
\usepackage{multirow}
\usepackage{amsmath, amsfonts, amssymb}
\usepackage{comment}
\usepackage{graphicx}
\usepackage{psfrag}
\usepackage{amsthm}
\usepackage[usenames,dvipsnames,svgnames,table]{xcolor}
\usepackage{enumerate}
\usepackage{arydshln}
\usepackage{soul}
 \usepackage{slashed}
\usepackage{subfig}
\usepackage{mathrsfs}
 \usepackage{a4wide}
  \usepackage{tikz}
  \usepackage{tikz-cd}
  \usetikzlibrary{shapes.geometric}
  \usepackage{tcolorbox}
  \usepackage{color}
  \definecolor{dark-gray}{gray}{0.20}
  \definecolor{gray}{gray}{0.30}
  \definecolor{light-gray}{gray}{0.80}
  \definecolor{dark-red}{rgb}{0.7,0,0}
  \definecolor{dark-green}{rgb}{0.1,0.4,0}
  \definecolor{dark-blue}{rgb}{0.3,0.3,0.7}
  \definecolor{light-blue}{rgb}{0.8,0.8,1}
      \definecolor{swamp}{RGB}{240, 199, 197}

  \usepackage{pifont}

\usepackage{newunicodechar} % To write the definition on the next line

\usepackage{setspace}

\newcommand{\be}{\begin{equation}}
\newcommand{\ee}{\end{equation}}
\newcommand{\eq}[1]{(\ref{#1})}

\def\be{\begin{equation}}
\def\ee{\end{equation}}
\def\bea{\begin{eqnarray}}
\def\eea{\end{eqnarray}}

\def\simleq{\; \raise0.3ex\hbox{$<$\kern-0.75em
      \raise-1.1ex\hbox{$\sim$}}\; }
   \def\simgeq{\; \raise0.3ex\hbox{$>$\kern-0.75em
      \raise-1.1ex\hbox{$\sim$}}\; }

\numberwithin{equation}{section}

\usepackage{jheppub}
\usepackage{hyperref}
\usepackage{cleveref}

\hypersetup{
	colorlinks=true,
	linkcolor=dark-blue,
	citecolor=dark-red,
	urlcolor=dark-green,
	linktoc=page
}

\theoremstyle{remark}

\crefname{appendix}{Appendix}{Appendices}

\title{\centering Black Holes as Probes of Moduli Space Geometry}

\author{Matilda Delgado$^1$,} \affiliation{$^1$ Instituto de F\'{i}sica Te\'{o}rica IFT-UAM/CSIC,
C/ Nicol\'{a}s Cabrera 13-15, Campus de Cantoblanco, 28049 Madrid, Spain}
\author{Miguel Montero$^2$,}\affiliation{$^2$Department of Physics, Harvard University, Cambridge, MA 02138, USA}
\author{Cumrun Vafa$^2$} 

\emailAdd{matilda.delgado@uam.es}
\emailAdd{mmontero@g.harvard.edu}
\emailAdd{vafa@g.harvard.edu}

\abstract{We argue that supersymmetric BPS states can act as efficient finite energy probes of the moduli space geometry thanks to the attractor mechanism. We focus on 4d $\mathcal{N}=2$ compactifications and capture aspects of the effective field theory near the attractor values in terms of physical quantities far away in moduli space. Furthermore, we illustrate how the standard distance in moduli space can be related asymptotically to the black hole mass.  We also compute a measure of the resolution with which BPS black holes of a given mass can distinguish far away points in the moduli space. The black hole probes may lead to a deeper understanding of the Swampland constraints on the geometry of the moduli space.}

\begin{document}
\hypersetup{pageanchor=false}
\makeatletter
\let\old@fpheader\@fpheader
\preprint{IFT-UAM/CSIC-22-151}

\makeatother

\maketitle

\hypersetup{
    pdftitle={Title},
    pdfauthor={Matilda Delgado, Miguel Montero, Cumrun Vafa},
    pdfsubject={String dualities, Anomalies, Cobordisms}
}

\newcommand{\remove}[1]{\textcolor{red}{\sout{#1}}}

\section{Introduction}

One of the basic features of String Theory compactifications is the ubiquity of moduli fields -- massless or light scalars that parametrize the internal compactification space and that control the masses and couplings of the theory. The geometry of moduli fields is relevant to the physical content of the theory and is captured by aspects of the Swampland Program including the Distance Conjecture \cite{Ooguri:2006in}. Given their relevance, it seems an important question how would we probe the global geometry of moduli fields in terms of physical data at one point in moduli space.

More concretely, suppose one has a theory with an exact or approximate moduli space, and we live in a vacuum where the moduli $\phi$ take some value $\phi_0$. One can study directly the physics of the vacuum at $\phi=\phi_0$ by means of scattering experiments, and even learn about the local geometry of the moduli space by studying these couplings. However it is, in general, very difficult to design an experiment to probe the physics at some value $\phi=\phi_1$ which is very far away from $\phi_0$.  It is precisely in faraway regions where interesting physics (such as decaying towers of states, emergence of perturbative string or decompactification limits, etc.) is supposed to take place. In most string theory literature, one is satisfied with studying the family of vacua parametrized by $\phi$, as well as the $\phi$ dependence of various observables such as masses and couplings. Yet this approach is somewhat unphysical: changing the vev of $\phi$ everywhere at once costs infinite energy, and once one starts considering configurations where $\phi$ only changes in a region of finite size, other challenges can appear.  One difficulty in designing setups that will probe large variations $\Delta \phi$ is that, when these are transplanckian, they will have a significant gravitational backreaction of their own, and whatever setup we consider is in danger of collapsing into a black hole. This was studied in \cite{Nicolis_2008}, as well as in the more recent series of papers \cite{Dolan:2017vmn,Draper:2019utz}, where it was pointed out that probing arbitrarily large $\Delta\phi$ is in principle possible in the effective field theory, but requires resources (masses, size of the laboratory\ldots ) exponentially large in $\Delta\phi$. Furthermore, the constructions described in \cite{Nicolis_2008} are not solving Einstein's equations, and so are regarded at best as interesting initial conditions, but they do not provide any concrete way to probe the faraway regions in moduli space.

The basic goal of this short note is to explain how the attractor mechanism allows one to overcome this challenge and do a form of ``black hole spectroscopy'', where properties of black holes at any one given point in moduli space can be used to probe the vacuum in faraway regions of the moduli space. Indeed, in a 4d $\mathcal{N}=2$ theory, thanks to the attractor mechanism (first constructed in \cite{Ferrara:1995ih} and further studied in \cite{Ferrara_1996a,Ferrara_1996b,Ferrara_1997,moore,Denef:2001xn}), the properties of the vector multiplet moduli space far away from any given vacuum can be studied reliably and in a robust way -- simply engineer a black hole such that the near-horizon values of all scalars $X_h^I$ differ significantly from those at infinity. The resulting geometry has the same asymptotics as the vacuum, but the near-horizon geometry constitutes a very long $AdS_2\times S^2$ geometry, where the scalars are stabilized at the attractor value. The two regions are joined by an intermediate throat in which the fields $X^I$ run. The size of the $S^2$ (or equivalently, the curvature of the $AdS_2$) are furthermore controlled by the total value of the black hole charge, which may be safely rescaled to arbitrarily large values without affecting the attractor solution. What this means is that one can, while keeping the attractor value fixed, engineer an $AdS_2\times S^2$ region where the size of the $S^2$ is arbitrarily large, and in which the physics looks locally like the vacuum solution on $\mathbb{R}^4$ with the attractor values of the moduli, thus achieving a concrete ``laboratory''  in which the asymptotic vacuum can be probed.  To make this picture concrete, we will show explicitly how the triple intersection numbers of the infinite distance limit, as well as the first subleading correction, can be encoded in term of mass and degeneracy of charged BPS states that can be physically measured in the asymptotic region.

If one has access to arbitrary mass/charge states, it becomes possible to study points in moduli space which are arbitrarily far away and with any desired precision. It is more interesting to study how the distance in moduli space and the resolution of the probing depend on the resources -- how well can we do if we have a maximum allowed mass, charge for the states. We study this question in a simple two-parameter family of black holes, finding agreement with the results in \cite{Nicolis_2008} that an exponential field range in moduli space are intimately related to the masses which trigger the flow. However, unlike in \cite{Nicolis_2008}, we have a concrete setup where transplanckian field ranges are attained in the context of a supersymmetric solution, in a time-independent way, thanks to the attractor mechanism.

The rest of the note includes a review of 4d $\mathcal{N}=2$ black holes and the attractor mechanism in Section \ref{sec:rev}, followed by the main application of black hole spectroscopy in Section \ref{sec:probing}, where we obtain the triple intersection numbers of an asymptotic limit in terms of degeneracy of states. Section \ref{sec:4} quantifies just how far can we go in probing the moduli space geometry for a given mass, and Section \ref{sec:5} explains how we quantify the resolution of points on moduli space using physical data at far away points.  Section \ref{sec:conclus} contains a few concluding remarks.

\section{Review of 4d \texorpdfstring{$\mathcal{N}=2$}{N=2} black hole solutions}\label{sec:rev}

We will start by reviewing some elements of Calabi-Yau three-fold $X$ compactifications of Type IIB string theory and their associated black hole solutions, which will be the core of this paper. The reader interested in further details is encouraged to check \cite{Greene:1996cy,LAbook2012}; here we will only describe the essentials of what we need. At low energies, the effective field theory describing a Calabi-Yau compactification is a four-dimensional $\mathcal{N}=2$ supergravity, coupled to $n_V= h^{1,2}$ abelian vector multiplets and $n_H= h^{1,1}+ 1$ hypermultiplets, where $h^{i,j}$ are the Hodge numbers of the Calabi-Yau three-fold $X$. The vector multiplet moduli space is a special K\"{a}hler manifold, and its scalars parametrize the complex structure of $X$. The dynamics of the hypermultiplets decouples completely from that of the vectors in the black holes we will consider, due to the 4d $\mathcal{N}=2$, so we will mostly ignore them in the following. 

As explained e.g. in \cite{Candelas:1990pi}, the intersection pairing in the middle cohomology of the Calabi-Yau defines a symplectic (antisymmetric) inner product. Constructing the complex structure moduli space comes down to choosing a symplectic basis $\{A^I, B_ I \}$ of 3-cycles in $H_3(X,\mathbb{Z})$ (and the corresponding basis of three forms $\{\alpha_ I, \beta^J \}$ of $H^3(X,\mathbb{Z})$), which we take to be orthonormal in the following sense: 
\begin{equation}\begin{gathered}
    \langle\alpha_ I ,\beta^J\rangle=-\langle\beta^J, \alpha_ I\rangle=\int_X \alpha_ I \wedge \beta^J = \delta_I ^J \\
    \int_{A^I}\alpha _ J = - \int _{B_J}\beta ^I = \delta ^I _ J \; ; \;\;\; \int_{A^I}\beta ^ J =  \int _{B_J}\alpha _I = 0\;,
\end{gathered}\end{equation}
where $\{I,J\}\in\{0,...,h^{2,1}\}$. Every Calabi-Yau manifold has a holomorphic (3,0)-form (see e.g. \cite{Greene:1996cy}) that can be decomposed as follows in terms of its A- and B-periods $\{X^I, F_J\}$:
\begin{equation}
    X^I = \int_ {A^I} \Omega_ 3 \;\;\;\;\; F_J = \int_{B_J} \;\Omega_3\;\;\;\longleftrightarrow \;\;\; \Omega_3=  X^I \alpha_ I - F_ J \beta^J \,.
\end{equation}
Performing a change $\Omega_3 \to e^f \Omega_3$ has no impact on the complex structure of X. In terms of the scalars $X^I$, this amounts to an overall re-scaling $X^I \to e^f X^I$ from which it is clear that only $h^{2,1}$ of these $h^{2,1}+1$ scalars are independent. The K\"{a}hler potential is given by (see. eg \cite{Polchinski:1998rr})
\begin{equation}\label{eq:Kahlerpotdef}
    \mathcal{K}=-ln \left(i \int_ X \Omega_3 \wedge \bar \Omega_ 3\right)=-ln\; i \left( \bar X ^I F_ I - X^I \bar F _ I  \right)\,.
\end{equation}
One can now see that a rescaling of $\Omega_3$ corresponds to a K\"{a}hler transformation in 4d $\mathcal{N}=2$ language: 
\begin{equation}
    \Omega_3 \to e^f \Omega_3 \;\;\;\;\; \mathcal{K}\to \mathcal{K}-f-\bar f\,.
\end{equation}
The fact that the complex structure of X is unchanged by a rescaling of $\Omega_3$ translates to the 4d $\mathcal{N}=2$ Lagrangian being invariant under K\"{a}hler transformations. One can therefore define a K\"{a}hler metric on the complex structure moduli space that is invariant under rescalings of $\Omega_3$ by using $\mathcal{K}$ as a K\"{a}hler potential. The metric obtained in this way on complex structure moduli space $g_{I\bar J}\sim \partial _ I \partial_ {\bar J} \mathcal{K}$ coincides with what can be read off of the kinetic term of the 4d Lagrangian for the complex structure moduli. One can choose a symplectic basis such that a single holomorphic function, the so-called prepotential  $F=F(X)$, encodes all the data of the topological theory. The B-periods can be reexpressed in terms of the prepotential as: 
\begin{equation}\label{eq:FIfromK}
   F_J(X)= \frac{\partial F(X)}{\partial X^J} .
\end{equation}

One can construct black hole solutions in the 4d $\mathcal{N}=2$ effective theory by wrapping D-branes on the various cycles of X (see eg. \cite{Denef_1999,Denef_2000,Denef:2001xn,moore}). These are generalizations of Reissner-Nordstr\"{o}m black holes, charged under the $n_V= h^{1,2}$ abelian vector multiplets. These black holes have the remarkable property that they are \textit{attractors} for the vector multiplet moduli. This means that these moduli, in general, run along the radial direction until they reach the black hole horizon where their value is entirely determined by the supersymmetric equations of motion, in what is known as the attractor mechanism \cite{Ferrara:1995ih}.  The attractor equations that describe this flow relate the charges of the black hole to the values of the moduli at the horizon. Throughout this work, we will use the attractor mechanism as a tool to map black hole thermodynamic properties to the complex structure moduli space. 

Let us now review the attractor mechanism of extremal 4d $\mathcal{N}=2$ black holes in more detail. In type IIB language, one constructs such black holes by wrapping D3 branes on a general 3-cycle $\mathcal{C}$ in X. Indeed, a black hole is identified by the decomposition of $\mathcal{C}$ onto the basis  $\{A^I, B_ I \}$ or equivalently by its corresponding electric and magnetic charges $\{p^I, q_ J\}$. Take $\Gamma$ to be the 3-form that is Poincar\'{e} dual of $\mathcal{C}$, then the corresponding splitting of magnetic and electric charges $\{p^I, q_ J\}$ is given by: 
\begin{equation}
   p^I  = \int_{A^I } \Gamma  \;\;\;\;\;   q_J = \int_{B_J} \;\Gamma \,.
\end{equation}

Consider a BPS solution charged under the 3-form $\Gamma$.  Then, the central charge of the black hole is given by: 
\begin{equation}
    Z=e^{\mathcal{K}/2}\int _ X \Omega_ 3\wedge \Gamma= e^{\mathcal{K}/2}(p^I F_ I-q_I X^I )\,.\label{ccharg}
\end{equation}
The attractor mechanism acts as a potential for the moduli and drives them to minimizing the central charge at the horizon of the black hole (note that the horizon values of the moduli will differ significantly, in general, from their values at spatial infinity) \cite{Ferrara_1997}. This minimization procedure leads to the attractor equations at the horizon, which relate the holomorphic periods to the charges of the black hole and can be written as follows: 
\begin{equation}\label{eq:attractoreqs1}
    \begin{gathered}
    p^I = \text{Re}\left[C_h X_h ^I \right]=C_h X_h ^I  + \bar C_h {\bar X_h} ^I \, , \\q_I = \text{Re}\left[C_h F_{h \,I} \right]=C_h F _{h\, I}  + \bar C_h {\bar F} _{h\,I} \, ,
    \end{gathered}
\end{equation}
where the ``h'' subscripts emphasize that these quantities are evaluated at the horizon and where we have introduced $C\equiv - 2 i \bar Z e^{\mathcal{K}/2}$. The vector multiplet moduli, which can be expressed in terms of the $X^I $, have arbitrary values infinitely far from the black hole, they vary along the radial direction and are fixed by the attractor equations at the horizon. Solving these equations for the periods at the horizon allows one to obtain the entropy of the black hole (equivalently, its area), which is expressed in terms of the central charge as: 
\begin{equation}\label{eq:entropy}S= \pi |Z_h|^2\,.\end{equation}
 One can also compute the ADM mass of the black hole, which turns out to be
\begin{equation}\label{eq:mass}
    M_{ADM}^2 = | Z_\infty |^2,\end{equation}
where we have introduced the subscript to emphasize that the ADM mass is obtained by evaluating the central charge $Z$, viewed as a function of the charges $p^I$, $q_I$ and the scalar values $X^I$ given in \eq{ccharg}, with the scalars $X^I$ taken to have their asymptotic values, i.e. evaluated at infinite distance from the black hole. In the particular case where the asymptotic and near-horizon values of the scalars coincide, \eq{eq:mass} and \eq{eq:attractoreqs1} agree: the attractor value of the mass is just given by the near-horizon dynamics. When they do not, the difference is due entirely to the running scalars outside of the horizon contributing to the mass. This follows from the attractor equations, which imply \cite{Ferrara_1997}
\begin{equation}|Z_\infty|^2-|Z_h|^2=\int_{-\infty}^0 d\tau\, e^{-U/2} \sqrt{g_{I\bar{J}} \frac{dt^I}{d\tau}\frac{dt^{\bar{J}}}{d\tau}}.\label{amass}\end{equation}
In this expression, $t^I\equiv X^I/X^0$ are the physical moduli, $e^{2U}$ is the time-component of the black hole metric, and $\tau$ is a certain parametrization of the radial coordinate in which the horizon sits at $\tau=-\infty$ and spatial infinity is at $\tau=0$.  Thus, we see that the difference in mass above the attractor value is just the backreaction of the running moduli.

Finally, we also note that the attractor equations and in particular the charges are invariant under K\"{a}hler transformations, which act on the periods and $C$ as follows: 
\begin{equation}\mathcal{K}\rightarrow \mathcal{K}- f -\bar f, \;\;\; C\rightarrow e^{-f} C, \;\;\; X^I \rightarrow e^{f} X^I, \;\;\; F \rightarrow e^{2f} F. \label{ktrans}\end{equation}
One can obtain the $ h^{2,1}$ physical, invariant, moduli $t^I$  by choosing special coordinates such as $X^I = t ^I X^0 $. Throughout the next sections we will be solving the attractor equations \eqref{eq:attractoreqs1} by choosing a constant $C_h$. Solving the attractor equations with different values of $C_h$ will generate a set of black hole solutions with the same attractor point in moduli space but different charges and masses. 

In the next section we will exploit the attractor equations in an attempt to map topological data of the Calabi-Yau moduli space to thermodynamic properties of black holes. 

\section{Probing the prepotential with large black holes}\label{sec:probing}
Armed with the attractor mechanism described in the previous Section, we will explain how it can be used to achieve a simple form of black hole moduli space spectroscopy, where we relate the properties of faraway points in moduli space to statistical, thermodynamic properties of large charge BPS states in a given vacuum. As described above, the attractor mechanism produces near-horizon $AdS_2\times S^2$ geometries where the value of the moduli are controlled by the attractor mechanism and can in general be very different from the asymptotic values of the moduli. For concreteness and simplicity, we will be interested in black holes that take the vector multiplet moduli to near-infinite distance limits in their moduli space. In these regions, the prepotential is constrained to take the well-known form: 
\begin{equation}\label{eq:cubicprep}F(X) = - D_{IJK}   \frac{X^I X^J X^K}{X^0}\,,\end{equation} where $C_{IJK}= 6 D_{IJK}$ are integers which, in the Calabi-Yau context, receive the interpretation of the triple intersection numbers of the mirror Calabi-Yau. But it is expected that this structure follows from general quantum gravity principles, even when a Calabi-Yau description is not present (see \cite{Cecotti:2020rjq}).

 With this, we see from \eqref{eq:FIfromK} that the attractor equations will relate the charges of the black hole directly to the $X^I$ at the horizon and the parameters $D_{IJK}$ (we take $C_h$ to be a constant at the horizon). Naturally, these charges will be very large since the moduli are reaching near-infinite values. Turning things around, solving these equations for the moduli at the horizon would allow us to express the entropy of the black hole \eqref{eq:entropy} in terms of the charges $\{p^I, q_ J\}$ and $D_{IJK}$. This would show that if one could measure the entropy and charge of one of these large black holes experimentally, it would be possible to deduce the values of the $D_{IJK}$. One would therefore recover topological data of the underlying Calabi-Yau from measuring black hole observables at a very different point in moduli space. Furthermore, quantities like electric and magnetic charges, or the degeneracy of charged BPS states (i.e., entropy of the black holes), are actual observables, which one could measure experimentally. 
 
We will just illustrate this method in the simplest example, and assume that we have a single vector multiplet $n_V= 1$. Then, there are just four periods, and from the prepotential \eqref{eq:cubicprep} we have  
\begin{equation}F_1= -3 D_{111} \frac{(X^1)^2}{X^0}\quad\text{and}\quad F_0 =  D_{111}\frac{(X^1)^3}{(X^0)^2}.\end{equation}
 One can set $C_h=1$ at the horizon by a K\"{a}hler transformation, and then the attractor equations are given by: 
\begin{align}\label{eqs:attracteqcub}
 p^0 &=  \;\text{Re}\left[ X ^0 \right] \;\;\;q_0 =  D_{111} \;\text{Re}\left[\frac{(X^1)^3}{(X^0)^2} \right]\nonumber\\
 p^1 &=  \;\text{Re}\left[X^1\right] \;\;\;q_1 = -3 D_{111} \; \text{Re}\left[\frac{(X^1)^2}{X^0}\right].
 \end{align}

Solving these equations yields the central charge at the horizon in terms of the $X^I $ fields, 
\begin{equation}|Z|^2=  \frac{ D_{111} |X^1 \bar{X^0}-X^0  \bar{X^1}|^3}{4 | X^0| ^4}\,.\end{equation}
Equivalently, one can solve \eqref{eqs:attracteqcub} for the periods and express the entropy in terms of the charges and $D_{111}$. For simplicity, we will assume that one of the charges vanishes ($p_0=0$), in which case we obtain the entropy as:  
\begin{equation}\label{eq:Scub}S=\pi |p^1|  \frac{ \sqrt{\vert q_1^2-12 D_{111} p^1 q_0}\vert }{\sqrt{3} }\,. \end{equation}
The argument of the square root is always positive if we pick charges such that the attractor equations have a solution. We emphasize that an expression such as \eq{eq:Scub} is anyway only expected to hold for very large charges, and in a one-parameter family of solutions such that the attractor values of the scalar are approaching the infinite distance limit in which \eq{eq:cubicprep} is approximately valid. One example of such a family can be parametrized as follows: in terms of the physical modulus $t = X^1/ X^0 $, take the charges that scale, in the $y \sim |t| \to \infty$ limit, as 
\begin{equation}\label{eq:chargelimit}
    \begin{gathered}
      \mathcal{Q}_\infty= \begin{cases}p^0= 0\\ p^1= N \\
       q_0= -y ^2N \\
       q_1 = 0 \end{cases}
    \end{gathered}\,.
\end{equation}
Here, $N$ is an overall rescaling of the charges, that does not affect the attractor value, but which will be important in a number of applications in what follows. 
Importantly, we have chosen a family of black holes whose charges solve \eqref{eqs:attracteqcub} but do not depend on $D_{111}$ explicitly. We are trying to encode $D_{111}$ in terms of observables such as charges and the degeneracy of BPS states, and therefore, choosing charges depending on $D_{111}$ would amount to assuming the answer. For the family \eq{eq:chargelimit}, the entropy formula simplifies to
\begin{equation}\label{eq:entropyfinal}S= 2 \pi \sqrt{ D_{111}}\, y N^2\,.\end{equation}
By counting the number of BPS states with such charges in a 4d $\mathcal{N}=2$ world, one could use this formula to obtain an experimental evaluation of the triple-intersection number $D_{111}$, and provides a direct link between moduli space properties and properties of the prepotential. This is an interesting result since one would not expect to be able to probe far away moduli data from measurements in the middle of moduli space, without any knowledge of the underlying compactification. Naturally, the formula \eqref{eq:entropyfinal} is to be taken as proof of concept that such a relation can be made. The exact expression will change if we consider a large black hole with charges that scale differently, or if one considers a different Calabi-Yau for the compactification.

A natural next step is to determine whether this framework can also be used to detect subleading corrections in the prepotential. To this effect, we will repeat the above analysis, now including the first correction to the prepotential as one moves slightly into the bulk of moduli space:
\begin{equation}\label{eq:correctedprep}F(X) = - D_{IJK}   \frac{X^I X^J X^K}{X^0} + d_ I X^I X^0. \end{equation}
In the Calabi-Yau case, we can relate $d_I$ to topological properties of the mirror Calabi-Yau via the formula
\begin{equation}\int _ {CY} c_2 \wedge \alpha _ I = 24 \,d_ I ,\end{equation}
 where $c_2$ and $\alpha_I$ are the second Chern class and the corresponding two-form of the mirror Calabi-Yau. Again, one can write the attractor equations in the simplified case where there is only one modulus, where they become
\begin{align}\label{eqs:attracteqcorr}
     p^0 &=  \;\text{Re}\left[ X ^0 \right]\;\;\;\;\;  q_0 =  D_{111} \;\text{Re}\left[\frac{(X^1)^3}{(X^0)^2} \right]+ d_1\;\text{Re}\left[X^1 \right] \\
     p^1 &=  \;\text{Re}\left[X^1\right]\;\;\;\;\;q_1 = -3 D_{111} \; \text{Re}\left[\frac{(X^1)^2}{X^0} \right]+ d_1\;\text{Re}\left[X^0\right]\,.\end{align}
One can solve these equations for the periods at the horizon and express the entropy in terms of the charges. For a black hole with a single vanishing charge $p^0=0$, one obtains: 
\begin{equation}S= \frac{\pi}{\sqrt{3}} | p^1|  \sqrt{12 d_1 D_{111} (p^1)^2-12 D_{111} p^1 q_0+(q^1)^2} \,.\end{equation}
It is easy to see that this reduces exactly to \eqref{eq:Scub} when $d_1$ is set to zero. One can evaluate the entropy of the large black hole with charges that scale as \eqref{eq:chargelimit} in the $y \sim |t| \to \infty$ limit and obtain:
\begin{equation}\label{eq:entropyfinalc2}S=2 \pi N^2 \sqrt{ D_{111}(y^2+ d_1)}\,. \end{equation}
Expanding this near $y\to \infty$, one obtains 
\begin{equation}\label{eq:entropyfinalc2lim}S=\pi N^2\Big[2 \sqrt{ D_{111}} y + d_1 \sqrt{ D_{111}} \frac{1}{y}  + \mathcal{O}(y^{-3})\Big].\end{equation}
 Having previously measured $D_{111}$ using \eqref{eq:entropyfinal} with extremely large black holes, one could measure deviations of this expression for slightly smaller black holes and obtain $d_1$ from \eqref{eq:entropyfinalc2}. 

An important point in all of the above is to note that we have been using the leading behavior of the black hole entropy.  This is valid for large $N$ and one expects to receive corrections suppressed by powers of $1/N^2$ \cite{Ooguri:2004zv}.  So for example if we want the $D_{111}$ term to be measurable, we need to ensure that 
\begin{equation}
    \sqrt{D_{111}}y\gtrsim\mathcal{O}(1/N^2)\, ,
\end{equation}
which is automatically satisfied in our case since $y\gg 1$ and $D_{111}$ is an integer. For the subleading term to be measurable we need to make sure 
\begin{equation}
   d_1 \sqrt{D_{111}}/y\gtrsim\mathcal{O}(1/N^2)\, ,
\end{equation}
which would be achievable if we pick $N\gtrsim\sqrt{y}$.  
In addition to polynomial corrections in prepotential, there are also exponential corrections:  
\begin{equation} \frac{F}{X_0^2}= -D_{IJK} t^I t^J t^K +d_I t^I + K_{\alpha_I} e^{-\alpha_I t^I} + \ldots,\end{equation}
where the coefficients $K_{\alpha_I}$ are quantized in the primitive directions \cite{Candelas:1990rm}.   In the one modulus example we have studied (where there is a single coefficient, $\alpha_1=1$), to get to the precision to be able to measure $K_{\alpha_I}$, we need to measure the degeneracy of large charge states, where
\begin{equation}
    \sqrt{D_{111}}\, N^2\gtrsim\mathcal{O}(y^3\, e^{2y})\, ,
\end{equation}
which is consistent with the intuition that measurement of exponentially small corrections require exponentially large charged BPS states.  As we will discuss in the next section $y$ itself is exponential in distance in moduli space, so this is a double exponential in terms of distance.

The above procedure could be refined indefinitely, recovering more and more information about the Calabi-Yau by measuring the sizes of smaller and smaller black holes. 
At each order in $y$, one can solve the attractor equations with the corrected prepotential and obtain the entropy in terms of the charges, the previously determined parameters and the new undetermined ones. Measuring the size of an appropriate selection of black holes will allow one to obtain the value of the undetermined parameters. The examples above describe the case with a single vector multiplet; in general, when $n_V>1$, one will need a multi-parameter family of black holes at each step. For instance, in the first step, one would need a large black hole with charges analogous to \eqref{eq:chargelimit} for each direction in moduli space in order to recover all of the triple intersection numbers $D_{IJK}$. 
Of course, this procedure becomes increasingly more complicated as we go further away from the controlled corners of the moduli space, though see \cite{Candelas:2021mwz}, where a similar iterative procedure was used to find generic solutions to the attractor equations at all orders, also incorporating instanton contributions. At low orders, this method is equivalent to our own. Nevertheless, the fact that such a procedure can be carried out in principle suggests that there is no obstruction in recovering the geometry of moduli space at any point, using physical measurements of charged BPS states at other points. However, the BPS degeneracy of states is a function of the attractor point alone, and it is not helpful in relating the asymptotic and attractor values of the moduli in a meaningful way. In the next Section we will address this question by considering the energetics of the BPS states and relating it to asymptotic distances in moduli space.

\section{Asymptotic black hole properties in moduli space}\label{sec:4}

As we saw in the previous Section, one can directly relate the prepotential, and thus, the usual moduli space metric, to degeneracy of BPS states. This suggests that it might be possible to obtain the full geometry of moduli space from other physical measurements.  
We now show that the attractor flow can correctly capture the notion of asymptotic distance near the boundaries of moduli space; namely, we will show that asymptotically in moduli space, the entropy of large BPS states and also their masses can be directly related to the distance in moduli space to their attractor points.

To do this, we once more consider black holes whose attractor point is near the boundary of moduli space, as in the previous Section.  
Using again the family of black holes with charges \eqref{eq:chargelimit}, the corresponding periods at the horizon are given by: 
\begin{equation}\begin{gathered}\text{Re}\left[X^0\right] = 0\;,\;\;\;\;\text{Im}\left[X^0\right] = -  N D_{111}^{1/2} y^{-1}\;,\\ \text{Re}\left[X^1\right] = N \;,\;\;\;\;\text{Im}\left[X^1\right] = 0\;.  \end{gathered}\end{equation}
From these expressions, it is straightforward to obtain the K\"{a}hler potential \eqref{eq:Kahlerpotdef} in this limit: 
\begin{equation}\mathcal{K}= -\log \left(8 N^2 \sqrt{D_{111}} y\right)\;.\end{equation}
The metric in moduli space is 
\begin{equation}ds^2 =2 \partial_{t}\partial_{ \bar t}K |dt|^2=  \frac{3}{2}\frac{|dt|^2}{\Im(t)^2}=\frac{3}{2}\frac{dy^2}{y^2}\; ,\end{equation}
so that the distance in moduli space is given by:
\begin{equation}d \sim  \sqrt{\frac{3}{2}} \log y\,.  \end{equation}
From the formula for the entropy \eqref{eq:entropyfinal} obtained for this set of charges, one immediately obtains
\begin{equation}\label{eq:SDeltat} S\sim N^2 e ^{\sqrt{\frac23} d}\,.\end{equation}
for this family of black hole solutions. A similar exponential relation holds asymptotically for the ADM mass: Using the general expression \eq{eq:mass}, one gets that
\begin{equation} M_{BPS}= e^{\mathcal{K_\infty}/2}\vert p_1 F_\infty^1-q_0 X_\infty^0\vert=N e^{\mathcal{K_\infty}/2}\vert F_\infty^1+y^2 X_\infty^0\vert,\end{equation}
where the subscript $_\infty$ in any quantity denotes its asymptotic values. For large $y$, the last term is leading, giving a dependence on the ADM mass that agrees parametrically with the entropy, and so 
\begin{equation}\label{eq:SDeltat2} M_{BPS}\sim N e ^{\sqrt{\frac83} d},\end{equation}
as well.

These expressions relate the mass of a large BPS state to the attractor point lying at a far away distance in moduli space.  In practice, it means that if one wants to probe the moduli space at a large distance $d$, one needs to create a massive BPS state with energy which is exponentially large in distance.  This is reminiscent of the work by Nicolis \cite{Nicolis_2008} where it was shown that, in a purely Newtonian setting, it was possible to construct setups with scalar sources that lead to arbitrarily large field ranges, with a size that grows exponentially on the field range. What we are finding is not only in agreement with this, but more broadly, with the findings of \cite{Dolan:2017vmn,Draper:2019utz}, which studied transplanckian field displacements in a variety of setups (including 4d dilatonic black holes). As proposed in \cite{Draper:2019utz}, there really seems to be a universal feature of quantum gravity that arbitrarily large field ranges can be probed at an exponential cost in physical resources. Other instances where one can see this include the extended objects of \cite{Lanza:2021udy,Angius:2022aeq} which probe infinite distances in moduli space; their tension goes exponentially with the distance. It would be very interesting to find the physical
mechanism underlying this phenomenon.

A scaling similar to \eqref{eq:SDeltat} was recovered in \cite{Bonnefoy_2020} in relation to the Black Hole Entropy Distance conjecture proposed in \cite{Lust:2019zwm}, a generalization of the ADC to black hole spacetimes. Both \cite{Lust:2019zwm,Bonnefoy_2020} proposed identifying the logarithm of the black hole entropy (the horizon area) with a notion of distance, encoded in the change of the metric when the flux changes one unit. The connection to \eq{eq:SDeltat2} is precisely that, when the quantized black hole charges change and the black hole area readjust, so do the vevs of the vector multiplet moduli, and the notion of distance using these or directly the metric as in \cite{Lust:2019zwm,Bonnefoy_2020} agree. In any case, equation \eq{eq:SDeltat} shows that the black holes provide a thermodynamic interpretation for the distance in moduli space, if only asymptotically.

Appendix \ref{app:A} discussed the precise realization of the general discussion in this paper in the context of a specific model, namely the mirror quintic $M$. The results agree with  \eq{eq:SDeltat}, as they should.  

\section{Resolution of the probing}\label{sec:5}
As we have seen, BPS states can serve as effective probes of far away regions of moduli space via their attractor geometry. However, when solving the attractor equations \eq{eqs:attracteqcub}, one needs to take into account Dirac quantization, which demands that the charges $p^I, q_I$ are quantized. In the Calabi-Yau picture, the quantization simply maps to the fact that the D3 branes that form the black holes we study must wrap an integer homology class. 

Dirac quantization implies that the picture of the moduli space provided by the black holes is not continuous; rather, it is naturally a mesh of points in the moduli space. These issues of quantization are however often ignored in the study of 4d $\mathcal{N}=2$ black holes, simply because of the K\"{a}hler transformation \eq{ktrans}. This transformation tells us that a homogeneous rescaling of the charges does not affect attractor values; as a result, one may simply scale the charges up, to very large values, achieving an arbitrarily dense mesh. While this is true, if one is constrained to finite resources (finite black hole mass, charge, or equivalently, entropy), the mesh allowed by Dirac quantization will be finite, leading to a finite ``resolution'' in the probing of moduli space. We will determine this resolution momentarily.

In more detail, consider the attractor equations near the boundary of moduli space \eqref{eqs:attracteqcub}. The general solution with non-vanishing charges $p^1>0,q^0<0$ is, asymptotically,
\begin{equation}X^1=p_1,\quad X_0=\sqrt{-\frac{p_1^3}{q_0}D_{111}},\quad t=\frac{X^1}{X^0}=\sqrt{-\frac{q_0}{p_1D_{111}}}.\label{modst}\end{equation}
We see that the attractor value of the physical modulus is insensitive to an overall rescaling of the charges.  As we make a small change in the charges, the value of $t$ in \eq{mods} changes, and we probe a nearby point of moduli space. The smallest such change that can take place, compatible with Dirac quantization, is changing $p_1$ by one unit while keeping $q_0$ constant. Under such a change, we obtain that the infinitesimal change in moduli space distance, $\delta d$, is given by
\begin{equation} \delta d=\sqrt{\frac32}\frac{\delta t}{t} =\sqrt{\frac38} \frac{1}{p_1},\label{eqd0}\end{equation}
where we have used the asymptotic form of the K\"{a}hler potential, $\mathcal{K}\sim-3\log t$. Using the asymptotic relation between the moduli space distance and the change in $t$,
\begin{equation} d\sim \sqrt{\frac32}\log t\quad\Rightarrow\quad t\sim e^{\sqrt{\frac23}d},\end{equation}
combined with \eq{modst}, one can rewrite \eq{eqd0} as
\begin{equation} \delta d=\sqrt{\frac38}\frac{t^2D_{111} }{\vert q_0\vert}\sim \sqrt{\frac38} \frac{D_{111}e^{\sqrt{\frac83}d}}{\vert q_0\vert},\label{e343}\end{equation}
in terms of the moduli space distance traversed by the black hole. Now, close to the infinite distance limit, $\vert t\vert\rightarrow\infty$, \eq{modst} tells us that $p_1$ is much smaller than $q_0$, and so the ADM mass 
\begin{equation} M_{BPS}=e^{\mathcal{K}_\infty/2}\left\vert p_{1} F^1_\infty-q_{0}X^0_\infty\right\vert,\label{mods}\end{equation}
can be approximated by the second term,
\begin{equation} M_{BPS}\approx  e^{\mathcal{K}_\infty/2}X^0_\infty  \vert q_0\vert.\end{equation}
This last equation allows us to replace $\vert q_0\vert$ by the black hole mass $M$ in \eq{e343}, yielding an expression
\begin{equation} \delta d=\sqrt{\frac38}{D_{111}e^{\mathcal{K}_\infty/2}X^0_\infty} \frac{e^{\sqrt{\frac83}d}}{M},\label{e344}\end{equation}
and finally, defining the resolution of the moduli space probing as the inverse spacing (in analogy with optics), we get
\begin{equation} r\equiv \frac{1}{\delta d} \sim \frac{M_{BPS}}{e^{\sqrt{\frac83}d}}=N,\label{res0}\end{equation}
where in the last equality we have used \eq{eq:SDeltat2}.
This equation gives us a notion of how the resolution in moduli space scales with the size of a large black hole whose attractor point is at a distance $d$ in moduli space. Taking $d$ to be a constant, we see that the resolution increases with the amount of energy (black hole mass) at one's disposal. This makes sense, as higher masses and bigger black holes naturally mean higher charges and so the ``mesh'' of points in moduli space becomes smaller.  Furthermore, keeping the mass of the black hole fixed, we see that the resolution will decrease exponentially as we try to explore farther points in moduli space. This fits naturally with previous discussions in \cite{Hamada:2021yxy} relating the Distance Conjecture to the Bekenstein bound and finiteness of quantum gravity amplitudes. The number of states that quantum gravity admits in a box should be finite, and bounded by the area of the box. This means that  it should not be possible to construct distinguishable states which probe infinite swaths of moduli space with arbitrarily large resolution in a box of given size. The resolution of this puzzle is, precisely, that the resolution drops quickly and makes far away points indistinguishable without increasing the size of the box. 

\section{Conclusions}\label{sec:conclus}
Although the physics of moduli spaces is arguably one of the most important aspects of the Swampland program and string compactifications, the question of how these moduli spaces could be probed in practice, if one was found, has received comparatively little attention.
In this short note we have shown how, in 4d $\mathcal{N}=2$ theories, asymptotic properties of moduli spaces are encoded in black hole solutions in a possibly very far away vacuum in moduli space, finding that quantitative features of the prepotential can be deduced from measurements of BPS degeneracies with appropriate charges. On top of this, the 4d $\mathcal{N}=2$ BPS states are able to reach arbitrarily large regions in moduli space, at a finite energy cost.

 We also studied the quality of the moduli space picture provided by the BPS black holes -- how far in moduli space can we go, and with which resolution, given finite resources. While this question is in general a complicated optimization problem, we studied a couple of one-parameter families of black holes, finding that both the distance goes logarithmically with the mass of the BPS black hole and that the resolution of the probing (which is finite due to charge quantization) depend linearly on the size of the black hole. This is in general agreement with the findings in \cite{Dolan:2017vmn,Draper:2019utz}, and shows that studying a transplanckian field range is possible in gravity, but takes an exponential amount of resources. It would be interesting to explore potential connections between this finding and more information-theoretic approaches to the Distance Conjecture recently put forth in \cite{Stout:2021ubb,Stout:2022phm}. 

 It is tantalizing that the distance conjecture is also an exponential relation between the mass of the tower and distance in moduli space. However, that involves the mass going exponentially down in distance, unlike the BPS mass that we need to probe such distances, which increases exponentially with distance.  It would be interesting to see if there is a relation between these two facts.
 
 One outstanding question is how to generalize our analysis to setups where supersymmetry is not protecting the answer, such as non-supersymmetric string theories or even questions involving hypermultiplets in 4d $\mathcal{N}=2$ theories, for which the attractor mechanism does not provide protection. Although it is likely that qualitative different ingredients are needed, we suspect that the basic point -- that black holes are appropriate probes of the moduli space -- is likely to apply, too.

Finally, the perspective we have taken in this manuscript is reminiscent of the moduli space holography picture of \cite{Grimm:2020cda,Grimm:2021ikg}. In that reference, it was proposed that the bulk of moduli space could be reconstructed from asymptotic data; in our setup, we have done the reverse, studying asymptotic regions from a bulk point in moduli space. And much as in the setup of \cite{Lanza:2021udy,Angius:2022aeq}, we have a one-to-one mapping between moduli space and physical space, sourced by the gradients of the fields. As these gradients can in principle be studied via ordinary holography, our perspective may help bridge the gap between these two disparate notions of holography.  

\vspace{0.5cm}
  
\textbf{Acknowledgements.}
We thank Max Wiesner and Angel Uranga for helpful discussions and comments on the manuscript.
 The work of MD is supported by the FPI grant no. FPI
SEV-2016-0597-19-3 from Spanish National Research Agency from the Ministry of Science and Innovation. MD also wishes to acknowledge the hospitality of the Department of Physics of Harvard University during the development of this work. The work of MM and CV is supported by a grant from the Simons Foundation (602883,CV) and by the NSF
grant PHY-2013858. The authors acknowledge the support of the Spanish Agencia Estatal de Investigacion through the grant “IFT Centro de Excelencia Severo Ochoa'' SEV-2016-0597.

\appendix

\section{Asymptotic mass formulae for the mirror quintic}\label{app:A}
In this Section we will particularize the general discussion in this paper to a specific model, namely the mirror quintic $M$. This is a Calabi-Yau threefold with hodge numbers $h^{1,1}= 101$ and $h^{2,1}= 1$. One can define it by considering the following quotient in $\mathbb{P}^4$:
\begin{equation}M = \left(\Sigma_ i z_i^5 - 5 \psi \Pi_ i z_i\right)/{\mathbb{Z}_5^3}\,.\end{equation}
The four periods of the mirror quintic were famously studied in \cite{Candelas:1990rm}. In particular, they can be combined into a period vector with respect to an integer symplectic basis $(A^i, B_j)$ of $H^3(M, \mathbb{Z})$, which in the large complex structure limit, $\psi \to \infty$, take the form \cite{Candelas:1990rm}:  
\begin{equation}\label{eq:QuinticPi}
    \Pi=\left(\begin{array}{c}F_0\\F_1 \\ X^0 \\ X^1 \\ \end{array}\right)\xrightarrow[]{\psi \to \infty}\Pi_{LCSL}=(\frac{2 \pi i}{5})^3\left(\begin{array}{c}\frac{5}{6}  t^3+ \frac{25}{12}t \\-\frac{5 }{2}t^2 -\frac{1}{2}t\\ 1 \\ t  \\ \end{array}\right),\;\;\text{with}\; \; t= - \frac{5}{2 \pi i}\log(5 \psi)\,.
\end{equation}
The K\"{a}hler potential at the large complex structure point is thus given by
\begin{equation}\label{kahlerpotLCS}
    e^{-\mathcal{K}}|_{LCSL}= \frac{32 \pi ^3  \log ^3(5 | \psi | )}{75}\;.
\end{equation}
Now, we will solve the attractor equations \eqref{eq:attractoreqs1} in a slightly different way than in the main text; rather than the choice of charges in \eq{eq:chargelimit}, we will use the choice of charges that exactly leads to the attractor values in \eq{eq:QuinticPi}, in a gauge where $C_h= N (\frac{2 \pi i}{5})^{-3}$ and writing $t= x+ i y$. One immediately obtains, in the limit $y \to \infty$: 
\begin{equation}\label{eq:quinticQ}
    \begin{gathered}
      \mathcal{Q}_\infty= \begin{cases}p^0 = N \\ p^1= Nx \\
       q_0= -\frac{5}{2 } N x y^2  \\
       q_1 = \frac{5}{2}Ny^2  \end{cases}.
    \end{gathered}
\end{equation}
From \eqref{kahlerpotLCS}, one obtains the distance in moduli space as $d= \sqrt{\frac{3}{2}}\log y$. Finally, we obtain the entropy from \eqref{eq:entropy} which, at leading order in $y$, is:
\begin{equation}S\sim N^2 y^3 \sim N^2 e^{\sqrt{6} d}\,. \end{equation}
This agrees with \eq{eq:SDeltat}, with a different exponent since we picked a different set of charges.
\bibliographystyle{JHEP}
\bibliography{references}

\providecommand{\href}[2]{#2}\begingroup\raggedright\begin{thebibliography}{10}

\bibitem{Ooguri:2006in}
H.~Ooguri and C.~Vafa, \emph{{On the Geometry of the String Landscape and the
  Swampland}},
  \href{https://doi.org/10.1016/j.nuclphysb.2006.10.033}{\emph{Nucl. Phys. B}
  {\bfseries 766} (2007) 21}
  [\href{https://arxiv.org/abs/hep-th/0605264}{{\ttfamily hep-th/0605264}}].

\bibitem{Nicolis_2008}
A.~Nicolis, \emph{On super-planckian fields at sub-planckian energies},
  \href{https://doi.org/10.1088/1126-6708/2008/07/023}{\emph{Journal of High
  Energy Physics} {\bfseries 2008} (2008) 023}.

\bibitem{Dolan:2017vmn}
M.J.~Dolan, P.~Draper, J.~Kozaczuk and H.~Patel, \emph{{Transplanckian
  Censorship and Global Cosmic Strings}},
  \href{https://doi.org/10.1007/JHEP04(2017)133}{\emph{JHEP} {\bfseries 04}
  (2017) 133} [\href{https://arxiv.org/abs/1701.05572}{{\ttfamily
  1701.05572}}].

\bibitem{Draper:2019utz}
P.~Draper and S.~Farkas, \emph{{Transplanckian Censorship and the Local
  Swampland Distance Conjecture}},
  \href{https://doi.org/10.1007/JHEP01(2020)133}{\emph{JHEP} {\bfseries 01}
  (2020) 133} [\href{https://arxiv.org/abs/1910.04804}{{\ttfamily
  1910.04804}}].

\bibitem{Ferrara:1995ih}
S.~Ferrara, R.~Kallosh and A.~Strominger, \emph{{N=2 extremal black holes}},
  \href{https://doi.org/10.1103/PhysRevD.52.R5412}{\emph{Phys. Rev. D}
  {\bfseries 52} (1995) R5412}
  [\href{https://arxiv.org/abs/hep-th/9508072}{{\ttfamily hep-th/9508072}}].

\bibitem{Ferrara_1996a}
S.~Ferrara and R.~Kallosh, \emph{Supersymmetry and attractors},
  \href{https://doi.org/10.1103/physrevd.54.1514}{\emph{Physical Review D}
  {\bfseries 54} (1996) 1514}.

\bibitem{Ferrara_1996b}
S.~Ferrara and R.~Kallosh, \emph{University of supersymmetric attractors},
  \href{https://doi.org/10.1103/physrevd.54.1525}{\emph{Physical Review D}
  {\bfseries 54} (1996) 1525}.

\bibitem{Ferrara_1997}
S.~Ferrara, G.W.~Gibbons and R.~Kallosh, \emph{Black holes and critical points
  in moduli space},
  \href{https://doi.org/10.1016/s0550-3213(97)00324-6}{\emph{Nuclear Physics B}
  {\bfseries 500} (1997) 75}.

\bibitem{moore}
G.~Moore, \emph{Arithmetic and attractors},
  \href{https://arxiv.org/abs/hep-th/9807087}{{\ttfamily hep-th/9807087}}.

\bibitem{Denef:2001xn}
F.~Denef, B.R.~Greene and M.~Raugas, \emph{{Split attractor flows and the
  spectrum of BPS D-branes on the quintic}},
  \href{https://doi.org/10.1088/1126-6708/2001/05/012}{\emph{JHEP} {\bfseries
  05} (2001) 012} [\href{https://arxiv.org/abs/hep-th/0101135}{{\ttfamily
  hep-th/0101135}}].

\bibitem{Greene:1996cy}
B.R.~Greene, \emph{{String theory on Calabi-Yau manifolds}},  in
  \emph{{Theoretical Advanced Study Institute in Elementary Particle Physics
  (TASI 96): Fields, Strings, and Duality}}, pp.~543--726, 6, 1996
  [\href{https://arxiv.org/abs/hep-th/9702155}{{\ttfamily hep-th/9702155}}].

\bibitem{LAbook2012}
L.E.~Ibáñez and A.M.~Uranga, \emph{String Theory and Particle Physics: An
  Introduction to String Phenomenology}, Cambridge University Press (2012),
  \href{https://doi.org/10.1017/CBO9781139018951}{10.1017/CBO9781139018951}.

\bibitem{Candelas:1990pi}
P.~Candelas and X.~de~la Ossa, \emph{{Moduli Space of {Calabi-Yau} Manifolds}},
  \href{https://doi.org/10.1016/0550-3213(91)90122-E}{\emph{Nucl. Phys. B}
  {\bfseries 355} (1991) 455}.

\bibitem{Polchinski:1998rr}
J.~Polchinski, \emph{{String theory. Vol. 2: Superstring theory and beyond}},
  Cambridge Monographs on Mathematical Physics, Cambridge University Press (12,
  2007),
  \href{https://doi.org/10.1017/CBO9780511618123}{10.1017/CBO9780511618123}.

\bibitem{Denef_1999}
F.~Denef, \emph{Attractors at weak gravity},
  \href{https://doi.org/10.1016/s0550-3213(99)00096-6}{\emph{Nuclear Physics B}
  {\bfseries 547} (1999) 201}.

\bibitem{Denef_2000}
F.~Denef, \emph{Supergravity flows and d-brane stability},
  \href{https://doi.org/10.1088/1126-6708/2000/08/050}{\emph{Journal of High
  Energy Physics} {\bfseries 2000} (2000) 050}.

\bibitem{Cecotti:2020rjq}
S.~Cecotti, \emph{{Special Geometry and the Swampland}},
  \href{https://doi.org/10.1007/JHEP09(2020)147}{\emph{JHEP} {\bfseries 09}
  (2020) 147} [\href{https://arxiv.org/abs/2004.06929}{{\ttfamily
  2004.06929}}].

\bibitem{Ooguri:2004zv}
H.~Ooguri, A.~Strominger and C.~Vafa, \emph{{Black hole attractors and the
  topological string}},
  \href{https://doi.org/10.1103/PhysRevD.70.106007}{\emph{Phys. Rev. D}
  {\bfseries 70} (2004) 106007}
  [\href{https://arxiv.org/abs/hep-th/0405146}{{\ttfamily hep-th/0405146}}].

\bibitem{Candelas:1990rm}
P.~Candelas, X.C.~De~La~Ossa, P.S.~Green and L.~Parkes, \emph{{A Pair of
  Calabi-Yau manifolds as an exactly soluble superconformal theory}},
  \href{https://doi.org/10.1016/0550-3213(91)90292-6}{\emph{Nucl. Phys. B}
  {\bfseries 359} (1991) 21}.

\bibitem{Candelas:2021mwz}
P.~Candelas, P.~Kuusela and J.~McGovern, \emph{{Attractors with large complex
  structure for one-parameter families of Calabi-Yau manifolds}},
  \href{https://doi.org/10.1007/JHEP11(2021)032}{\emph{JHEP} {\bfseries 11}
  (2021) 032} [\href{https://arxiv.org/abs/2104.02718}{{\ttfamily
  2104.02718}}].

\bibitem{Lanza:2021udy}
S.~Lanza, F.~Marchesano, L.~Martucci and I.~Valenzuela, \emph{{The EFT stringy
  viewpoint on large distances}},
  \href{https://doi.org/10.1007/JHEP09(2021)197}{\emph{JHEP} {\bfseries 09}
  (2021) 197} [\href{https://arxiv.org/abs/2104.05726}{{\ttfamily
  2104.05726}}].

\bibitem{Angius:2022aeq}
R.~Angius, J.~Calder\'on-Infante, M.~Delgado, J.~Huertas and A.M.~Uranga,
  \emph{{At the end of the world: Local Dynamical Cobordism}},
  \href{https://doi.org/10.1007/JHEP06(2022)142}{\emph{JHEP} {\bfseries 06}
  (2022) 142} [\href{https://arxiv.org/abs/2203.11240}{{\ttfamily
  2203.11240}}].

\bibitem{Bonnefoy_2020}
Q.~Bonnefoy, L.~Ciambelli, D.~Lüst and S.~Lüst, \emph{Infinite black hole
  entropies at infinite distances and tower of states},
  \href{https://doi.org/10.1016/j.nuclphysb.2020.115112}{\emph{Nuclear Physics
  B} {\bfseries 958} (2020) 115112}.

\bibitem{Lust:2019zwm}
D.~L\"ust, E.~Palti and C.~Vafa, \emph{{AdS and the Swampland}},
  \href{https://doi.org/10.1016/j.physletb.2019.134867}{\emph{Phys. Lett. B}
  {\bfseries 797} (2019) 134867}
  [\href{https://arxiv.org/abs/1906.05225}{{\ttfamily 1906.05225}}].

\bibitem{Hamada:2021yxy}
Y.~Hamada, M.~Montero, C.~Vafa and I.~Valenzuela, \emph{{Finiteness and the
  swampland}}, \href{https://doi.org/10.1088/1751-8121/ac6404}{\emph{J. Phys.
  A} {\bfseries 55} (2022) 224005}
  [\href{https://arxiv.org/abs/2111.00015}{{\ttfamily 2111.00015}}].

\bibitem{Stout:2021ubb}
J.~Stout, \emph{{Infinite Distance Limits and Information Theory}},
  \href{https://arxiv.org/abs/2106.11313}{{\ttfamily 2106.11313}}.

\bibitem{Stout:2022phm}
J.~Stout, \emph{{Infinite Distances and Factorization}},
  \href{https://arxiv.org/abs/2208.08444}{{\ttfamily 2208.08444}}.

\bibitem{Grimm:2020cda}
T.W.~Grimm, \emph{{Moduli space holography and the finiteness of flux vacua}},
  \href{https://doi.org/10.1007/JHEP10(2021)153}{\emph{JHEP} {\bfseries 10}
  (2021) 153} [\href{https://arxiv.org/abs/2010.15838}{{\ttfamily
  2010.15838}}].

\bibitem{Grimm:2021ikg}
T.W.~Grimm, J.~Monnee and D.~van~de Heisteeg, \emph{{Bulk reconstruction in
  moduli space holography}},
  \href{https://doi.org/10.1007/JHEP05(2022)010}{\emph{JHEP} {\bfseries 05}
  (2022) 010} [\href{https://arxiv.org/abs/2103.12746}{{\ttfamily
  2103.12746}}].

\end{thebibliography}\endgroup
\end{document}